\documentclass{ws-p8-50x6-00}

\begin{document}

\title{Recent BCP progress in Taiwan}

\author{Hsiang-nan Li}

\address{Department of Physics, National Cheng-Kung University, 
Tainan, Taiwan 701, Republic of China}

\address{Physics Division, National Center for Theoretical Sciences,
Hsinchu, Taiwan 300, Republic of China}


\maketitle

\abstracts{
I review theoretical progresses on $B$ physics and CP violation which were
made in Taiwan recently. I concentrate on the approaches to exclusive $B$
meson decays based on factorization assumption, $SU(3)$ symmetry,
perturbative QCD factorization theorem, QCD factorization, and
light-front QCD formalism.
}

\section{Introduction}

The collaboration on $B$ physics and CP violation (BCP) is one of the most
active groups in Taiwan. In this talk I will briefly review theoretical
progresses on BCP, which were made in Taiwan recently. For experimental
review, please refer to Dr. Wang's talk in this workshop. I will
concentrate on five approaches to exclusive $B$ meson decays based on
factorization assumption (FA), $SU(3)$ symmetry, perturbative QCD (PQCD)
factorization theorem, QCD factorization (QCDF), and light-front QCD
(LFQCD) formalism. Abundant results of exclusive $B$ meson decays have
been produced and important dynamics has been explored.

The effective Hamiltonian for $b$ quark decays, for example, the $b\to s$
transition, is given by\cite{REVIEW}
\begin{eqnarray}
H_{\rm eff}={G_F\over\sqrt{2}}
\sum_{q=u,c}V_q\left[C_1(\mu)O_1^{(q)}(\mu)
+C_2(\mu)O_2^{(q)}(\mu)+
\sum_{i=3}^{10}C_i(\mu)O_i(\mu)\right]\;,
\label{hbk}
\end{eqnarray}
with the Cabibbo-Kobayashi-Maskawa (CKM) matrix elements
$V_q=V^*_{qs}V_{qb}$. Using the unitarity condition, the CKM matrix
elements for the penguin operators $O_3$-$O_{10}$ can also be expressed
as $V_u+V_c=-V_t$. The definitions of the operators $O_i$ are referred
to\cite{REVIEW}.

According to the Wolfenstein parametrization, the CKM matrix upto
${\cal O}(\lambda^{3})$ is written as
\begin{eqnarray}
\left(\matrix{V_{ud} & V_{us} & V_{ub} \cr
              V_{cd} & V_{cs} & V_{cb} \cr
              V_{td} & V_{ts} & V_{tb} \cr}\right)
=\left(\matrix{ 1 - { \lambda^2 \over 2 } & \lambda &
A \lambda^3(\rho - i \eta)\cr
- \lambda & 1 - { \lambda^2 \over 2 } & A \lambda^2\cr
A \lambda^3(1-\rho-i\eta) & -A \lambda^2 & 1 \cr}\right)\;,
\label{ckm}
\end{eqnarray}
with the parameters\cite{LEP} $\lambda = 0.2196 \pm 0.0023$,
$A = 0.819 \pm 0.035$, and $R_b \equiv\sqrt{{\rho}^2  + {\eta}^2}
= 0.41 \pm 0.07$. 
The unitarity angle $\phi_3$ is defined via
\begin{equation}
V_{ub}=|V_{ub}|\exp(-i\phi_3)\;.
\end{equation} 

One of the important missions of $B$ fatories is to determine the angles
$\phi_1$, $\phi_2$ and $\phi_3$. The angle $\phi_1$ can be extracted from
the CP asymmetry in the $B\to J/\psi K_S$ decays, which arises from the
$B$-$\bar B$ mixing. Due to similar mechanism of CP asymmetry, the decays
$B^0\to \pi^+\pi^-$ are appropriate for the extraction of the angle
$\phi_2$. However, these modes contain penguin contributions such that
the extraction suffers large uncertainty. It has been proposed
that the angle $\phi_3$ can be determined from the decays $B\to K\pi$,
$\pi\pi$\cite{GRL,FM,NR,BF}. Contributions to these modes involve 
interference between penguin and tree amplitudes, and their analyses
rely heavily on theoretical formulations.

\section{Factorization Assumption}

The conventional approach to exclusive nonleptonic $B$ meson decays is
based on FA\cite{BSW}, in which nonfactorizable and annihilation
contributions are neglected and final-state-interaction (FSI) effects are
assumed to be absent. Factorizable contributions are expressed as
products of Wilson coefficients and hadronic form factors, which are then
parametrized by models. With these approximations, the FA approach is
simple and provides qualitative estimation of various decay branching
ratios.

To extract the angle $\phi_3$, we consider the ratios $R$ and $R_\pi$
defined by
\begin{eqnarray}
R=\frac{{\rm Br}(B_d^0\to K^\pm\pi^\mp)}{{\rm Br}
(B^\pm\to K^0\pi^\pm)}\;,& &
R_\pi=\frac{{\rm Br}(B_d^0 \to  K^\pm \pi^\mp)}
{{\rm Br}(B_d^0 \to \pi^\pm \pi^\mp)}\;,
\end{eqnarray}
where ${\rm Br}(B_d^0\to K^\pm\pi^\mp)$ represents the CP average of the
branching ratios ${\rm Br}(B_d^0\to K^+\pi^-)$ and
${\rm Br}({\bar B}_d^0\to K^-\pi^+)$, and the definition of
${\rm Br}(B^\pm\to K^0\pi^\pm)$ is similar. It has been shown that the
data $R\sim 1$ imply $\phi_3 \sim 90^o$\cite{WS}.

To explain the data of $R_\pi\sim 4$, a large angle $\phi_3\sim 130^o$
must be postulated\cite{WS}. It is easy to observe from Eqs.~(\ref{ckm})
and (\ref{ckp}) that the products of the CKM matrix elements
$V^*_{us}V_{ub}$ and $V^*_{ud}V_{ub}$ have the same weak phase, and that
the real parts of $V^*_{ts}V_{tb}$ and $V^*_{td}V_{ub}$ are opposite in
sign. That is, the tree-penguin interference in the decays $B\to K\pi$
and $B\to\pi\pi$ is anti-correlated. A $\phi_3 > 90^o$ then leads to
constructive interference between the tree and penguin contributions in
$B\to K\pi$, and a large $R_\pi$. The determination $\phi_3\sim 114^o$
from the global fit to charmless $B$ meson decays\cite{WS}, located
between the two extreme cases $90^o$ and $130^o$, is then understood. On
the other hand, a large $B\to\rho$ form factor $A^{B\rho}_0\sim 0.48$ has
been extracted, which accounts for the large $B\to\rho\pi$ branching
ratios. In the modified FA approach with an effective number of colors
$N_c^{\rm eff}$, a large unitarity angle $\phi_3\sim 105^o$ is also
concluded\cite{HYCheng}.

The above $\phi_3$, located in the second quadrant, contradict the
extraction from other measurements, such as $B_s$-mixing. The best fit to
experimental data of semileptonic $B$ meson decays, $B$-$\bar B$ mixing,
and $\epsilon_K$ indicates that $\phi_3$ is located in the first
quadrant. An improvement of FA has been considered, in which possible
strong phases produced via FSI are introduced as arbitrary
parameters\cite{HY}. Performing the best fit to data, a large $\phi_3$ is
still required and strong phases are found to be large, which generate
significant CP asymmetries in $K\pi$ and $\pi\pi$ modes.

\section{$SU(3)$ Symmetry}

The model-dependent determination of the angle $\phi_3$ from FA seems not
to be satisfactory. A more model-independent approach based on $SU(3)$
symmetry has been proposed\cite{SH}, in which the light quarks $u$, $d$
and $s$ form a $SU(3)$ triplet, while the heavy quarks $c$, $b$, and $t$
form $SU(3)$ singlets. According to the above assumption, the $B$ mesons
$B_u$, $B_d$, and $B_s$ form a $SU(3)$ triplet at the hadronic level.
Pseudo-scalar mesons $P$ and vector mesons $V$ also possess definite
$SU(3)$ structures,
\begin{eqnarray}
M^i_j&=&\left(\matrix{
\frac{\pi^0}{\sqrt{2}}-\frac{\eta_8}{\sqrt{6}}+\frac{\eta_1}{\sqrt{3}}
 & \pi^- & K^- \cr
\pi^+ & -\frac{\pi^0}{\sqrt{2}}-\frac{\eta_8}{\sqrt{6}}
+\frac{\eta_1}{\sqrt{3}} & {\bar K}^0 \cr
K^+ & K^0 &-\frac{2\eta_8}{\sqrt{6}}+\frac{\eta_1}{\sqrt{3}} \cr}\right)\;,
\\
V^i_j&=&\left(\matrix{
\frac{\rho^0}{\sqrt{2}}+\frac{\omega}{\sqrt{2}} & \rho^- & K^{*-} \cr
\rho^+ & -\frac{\rho^0}{\sqrt{2}}+\frac{\omega}{\sqrt{2}} &
{\bar K}^{*0} \cr
K^{*+} & K^{*0} & \phi \cr}\right)\;.
\end{eqnarray}

Similarly, the four-fermion operators in the weak Hamiltonian can be
decomposed into operators with definite $SU(3)$ structures. For example,
the penguin operators $O_{3-6}$ are labelled as $\bar 3$ states,
\begin{eqnarray}
{\bar q}b(\bar u u+\bar d d+\bar s s)\leftrightarrow \bar 3\;,
\label{pd}
\end{eqnarray}
since $\bar qb$ forms a triplet $\bar 3$ and $\bar u u+\bar d d+\bar s s$
froms a singlet. The operators $O_{1,2}$ are written as
\begin{eqnarray}
{\bar q}u \bar u b &\leftrightarrow& \bar 3\times 3\times \bar 3
=\bar 3+\bar 3+6 +\bar{15}\;.
\label{td}
\end{eqnarray}
Following Eqs.~(\ref{pd}) and (\ref{td}), the effective Hamiltonian in
Eq.~(\ref{hbk}) is decomposed into operators carrying different $SU(3)$
structures, such as $H(\bar 3)$, $H(6)$ and $H(\bar{15})$, whose
coefficients are the linear combinations of the Wilson coefficients.

Employing the above results, we formulate various decay amlitudes of the
tree and annihilation topologies for $B\to PP$ modes. For example, the
parameter $C_{\bar 3}$ associated with the contraction
$B_i M^i_k M^k_j H(\bar 3)^j$ represents a tree amplitude. The parameter
$A_{\bar 3}$ associated with the contraction $B_i H(\bar 3)^i M^k_lM^l_k$
represents an annihilation amplitude. Collecting all expressions for the
branching ratios and the CP asymmetries of the $B\to \pi\pi$, $K\pi$ and
$KK$ modes, there are totally 13 free parameters. This number is too big
for a global analysis of currently available data. As an approximation,
annihilation contributions are neglected. 8 parameters, the absolute
values of $C_{\bar 3}^T$, $C_{\bar 3}^P$, $C_6^T$ and $C_{\bar 15}^T$,
the phases $\delta_{\bar 3}^P$, $\delta_6^T$, and $\delta_{\bar 15}^T$,
and the CKM phase $\phi_3$, are then left, where $T$ ($P$) denotes the
tree operators $O_{1,2}$ (the penguin operators $O_{3-10}$).

The best fit to data gives
\begin{eqnarray}
& &\phi_3=70^o\;,\;\;\; \rho=0.17\;,\;\;\; \eta=0.37\;,
\nonumber\\
& &C_{\bar 3}^T=0.28\;,\;\;\; C_{\bar 3}^P=0.14\;,\;\;\;
C_6^T=0.33\;, \;\;\; C_{\bar 15}^T=0.14\;,
\nonumber\\
& &\delta_{\bar 3}^P=12^o\;,\;\;\; \delta_6^T=6^o\;,\;\;\;
\delta_{\bar 15}^T=74^o\;.
\end{eqnarray}
If the $SU(3)$ symmetry breaking effect from $f_K/f_\pi\not = 1$ is taken
into account, $f_K$ ($f_\pi$) being the kaon (pion) decay constant,
the results are shifted only a bit.
Hence, there is no indication that $\phi_3$ should be located in the
second quadrant. Certainly, the allowed range of the above parameters
is still large.

\section{Perturbative QCD}

It has been shown that the decay amplitudes and strong phases
discussed in the previous sections can be evaluated in the PQCD framework,
and that it is possible to extract $\phi_3$ from the
$B\to K\pi$ data\cite{KLS}. According to PQCD factorization theorem, a
$B$ meson decay amplitude is expressed as convolution of a hard $b$ quark
decay amplitude with meson wave functions. A meson wave function,
absorbing nonperturbative dynamics of a QCD process, is not calculable,
while a hard amplitude is.

In perturbation theory nonperturbative dynamics is reflected by infrared
divergences in radiative corrections. It has been proved to all orders
that these infrared divergences can be separated and absorbed into meson
wave functions\cite{L6}. A formal definition of wave functions as matrix
elements of nonlocal operators has been constructed, which, if evaluated
perturbatively, reproduces the infrared divergences. The gauge invariance
of the above factorization has been proved in\cite{CLY}. A meson wave
function must be determined by nonperturbative means, such as lattice
gauge theory and QCD sum rules, or extracted from experimental data. A
salient feature of PQCD factorization theorem is the universality of
nonperturbative wave functions. Because of universality, a $B$ meson wave
function extracted from some decay modes can be employed to make
predictions for other modes. This is the reason PQCD factorization
theorem possesses a predictive power.

In the practical calculation small parton transverse momenta $k_T$ are
included\cite{LS}, which are essential for smearing the end-point
singularities from small momentum fractions\cite{KLS}. Because of the
inclusion of $k_T$, double logarithms $\ln^2(Pb)$ are generated from the
overlap of collinear and soft enhancements in radiative corrections to
meson wave functions, where $P$ denotes the dominant light-cone component
of a meson momentum and $b$ is a variable conjugate to $k_T$. The
resummation of these double logarithms leads to a Sudakov form factor
$\exp[-s(P,b)]$\cite{CS,BS}, which suppresses the long-distance
contributions in the large $b$ region, and vanishes as $b=1/\Lambda$,
$\Lambda\equiv \Lambda_{\rm QCD}$ being the QCD scale. This suppression
guarantees the applicability of PQCD to exclusive decays around the
energy scale of the $b$ quark mass\cite{LY1}.

The hard amplitude contains all possible Feynman diagrams\cite{CL,YL},
such as factorizable diagrams, where hard gluons attach the valence
quarks in the same meson, and nonfactorizable diagrams, where hard gluons
attach the valence quarks in different mesons. The annihilation topology
is also included, and classified into factorizable or nonfactorizable
one. Therefore, FA for two-body $B$ meson decays is not necessary.
It has been shown that factorizable annihilation contributions are in
fact important, and give large strong phases in PQCD\cite{KLS}.

We emphasize that the hard amplitude is characterized by the virtuality
of internal particles, $t\sim\sqrt{\bar\Lambda M_B}\sim 1.5$ GeV,
$\bar\Lambda=M_B-m_b$. The RG evolution of the Wilson coefficients
$C_{4,6}(t)$ dramatically increase as $t<M_B/2$, such that penguin
contributions are enhanced\cite{KLS,LUY}. With this penguin enhancement,
the observed branching ratios of the $B\to K\pi$ and $B\to\pi\pi$ decays
can be explained in PQCD using a smaller angle $\phi_3=90^o$. That is,
the data of $R_\pi$ do not demand large $\phi_3$. Such a dynamical
enhancement of penguin contributions does not exist in the FA approach.

Our predictions for the branching ratio of each $K\pi$ mode corresponding
to $\phi_3=90^o$\cite{KLS},
\begin{eqnarray}
& &{\rm Br}(B^+\to K^0\pi^+)=21.72\times 10^{-6}\;,\;\;\;
{\rm Br}(B^-\to \bar{K}^0\pi^-)=21.25\times 10^{-6}\;,
\nonumber\\
& &{\rm Br}(B_d^0\to K^+\pi^-)=24.19\times 10^{-6} \;,\;\;\;
{\rm Br}({\bar B}_d^0\to K^-\pi^+)=16.84\times 10^{-6}\;,
\nonumber\\
& & {\rm Br}(B^+\to K^+\pi^0)= 14.44\times 10^{-6}\;,\;\;\;
{\rm Br}(B^-\to K^-\pi^0)= 10.65\times 10^{-6}\;,
\nonumber\\
& &{\rm Br}(B_d^0\to K^0\pi^0)= 11.23\times 10^{-6} \;,\;\;\;
{\rm Br}({\bar B}_d^0\to \bar{K}^0\pi^0)= 11.84\times 10^{-6}\;,
\label{pqp}
\end{eqnarray}
are consistent with the CLEO data\cite{YK},
\begin{eqnarray}
& &{\rm Br}(B^\pm\to K^0\pi^\pm)
=(18.2^{+4.6}_{-4.0}\pm 1.6)\times 10^{-6}\;,
\nonumber\\
& &{\rm Br}(B_d^0\to K^\pm\pi^\mp)
=(17.2^{+2.5}_{-2.4}\pm 1.2)\times 10^{-6} \;,
\nonumber\\
& &{\rm Br}(B^\pm\to K^\pm\pi^0)=(11.6^{+3.0+1.4}_{-2.7-1.3})
\times 10^{-6}\;,
\nonumber\\
& &{\rm Br}(B_d^0\to K^0\pi^0)=(14.6^{+5.9+2.4}_{-5.1-3.3})
\times 10^{-6} \;,
\nonumber\\
& &A_{CP}(B_d^0\to K^\pm\pi^\mp)=-0.04\pm 0.16\;,
\nonumber\\
& &A_{CP}(B^\pm\to K^0\pi^\pm)=0.17\pm 0.24\;.
\label{cld}
\end{eqnarray}
In the above expressions $B(B_d^0\to K^\pm\pi^\mp)$
represents the CP average of the branching ratios
$B(B_d^0\to K^+\pi^-)$ and $B({\bar B}_d^0\to K^-\pi^+)$.

\section{QCD Factorization}

Recently, Beneke, Buchalla, Neubert, and Sachrajda proposed the QCDF
formalism for two-body nonleptonic $B$ meson decays\cite{BBNS}. They
claimed that factorizable contributions, for example, the form factor
$F^{B\pi}$ in the $B\to\pi\pi$ decays, are not calculable in PQCD, but
nonfactorzable contributions are in the heavy quark limit. Hence, the
former are treated in the same way as FA, and expressed as products of
Wilson coefficients and $F^{B\pi}$. The latter, calculated as in the PQCD
approach, are written as the convolutions of hard amplitudes with three
$(B,\pi,\pi)$ meson wave functions. Annihilation diagrams are neglected
as in FA, but can be included as $1/M_B$ correction. Values of form
factors at maximal recoil and nonperturbative meson wave functions are
treated as free parameters.

Here I mention some essential differences between the QCDF and PQCD
approaches. For more detailed comparisions, refer to\cite{KL}. Because
of the neglect of annihilation diagrams in QCDF, strong phases and CP
asymmetries are much smaller than those predicted in PQCD. In
QCDF the leading-order diagrams are those that contain vertex corrections
to the four-fermion operators. These diagrams, however, appear at the
next-to-leading order in PQCD. This difference implies different
characteristic scales in the two approaches: the former is characterized
by the $b$ quark mass $m_b$, while the latter is characterized by the
virtuality $t$ of internal particles, which leads to the penguin
enhancement emphasized above. Without penguin enhancement, a large
$\phi_3$ is still necessary to account for the large ratio
$R_\pi$\cite{Du}.

The $B\to\phi K$ decays have been analyzed in the QCDF
formalism\cite{CY,HMS}, and branching ratios much smaller than
experimental data have been obtained. Since these modes are dominated by
penguin contributions, the penguin enhancement may be crucial for
explaining the data.

\section{Light-front QCD}

The evaluation of a form factor is simple in the LFQCD formalism, which
is written as an overlap integral of initial- and final-state meson wave
functions\cite{CCHZ}. Various form factors have been computed, such as
the $B\to \pi$, $\rho$, $K$ and $K^*$ form factors \cite{CHZ} and the
$\Lambda_b\to\Lambda$ form factors\cite{CG}. The results have been
employed to predict the decay spectra of the $B\to K\mu\mu (\tau\tau)$
and $\Lambda_b\to \Lambda\mu\mu (\tau\tau)$ modes. This formalism has
been also applied to the radiative leptonic $B$ meson decays
$B\to l^+ l^-\gamma$\cite{GLZ}. These predictions can be compared with
data in the future.

\section{Conclusion}

In this talk I have briefly summarized the theoretical progresses on
exclusive $B$ meson decays, which were made by Taiwan BCP community
recently. With the active collaboration, more progresses are expected in
the near future.

\section*{Acknowledgments}
This work was supported in part by the National Science Council
of R.O.C. under the Grant No. NSC-89-2112-M-006-033,
and in part by Grant-in Aid for Special Project Research
(Physics of CP Violation) and by Grant-in Aid for Scientific Exchange
from Ministry of Education, Science and Culture of Japan.

\end{document}